\documentclass[journal=jacsat,manuscript=article]{achemso}

\usepackage[version=3]{mhchem} % Formula subscripts using \ce{}
\usepackage[utf8]{inputenc}
\usepackage{graphicx}% Include figure files
\usepackage{dcolumn}% Align table columns on decimal point
\usepackage{bm}% bold math
\usepackage{blindtext}
\usepackage{fancyvrb}
\usepackage[normalem]{ulem}
\usepackage{arydshln}

\usepackage[justification=centering]{caption}
\newcommand\Tstrut{\rule{0pt}{2.9ex}}         % "top" strut
\newcommand\Bstrut{\rule[-1.2ex]{0pt}{0pt}}   % "bottom" strut
\newcommand\TBstrut{\Tstrut\Bstrut}           % "top and bottom" stru

\usepackage{tikz}
\usepackage{ulem}
\usetikzlibrary{shapes,arrows}
\usepackage{multirow}
\usepackage{amsmath}

\mciteErrorOnUnknownfalse
\author{Nastasia Mauger}
\affiliation{Sorbonne Université, Laboratoire de Chimie Théorique, UMR 7616 CNRS, 75005 Paris, France}
\alsoaffiliation{Department of Chemistry, University of Pittsburgh, Pittsburgh, Pennsylvania 15260, United States}
\email{nastasia.mauger@pitt.edu}

\author{Thomas Plé}

\affiliation{Sorbonne Université, Laboratoire de Chimie Théorique, UMR 7616 CNRS, 75005 Paris, France}
\author{Louis Lagardère}
\affiliation{Sorbonne Université, Laboratoire de Chimie Théorique, UMR 7616 CNRS, 75005 Paris, France}

\author{Simon Huppert}
\affiliation{Sorbonne Universit\'e, Institut des NanoSciences de Paris, UMR 7588 CNRS, 75005 Paris, France}
\email{simon.huppert@sorbonne-universite.fr}
\author{Jean-Philip Piquemal}
\affiliation{Sorbonne Université, Laboratoire de Chimie Théorique, UMR 7616 CNRS, 75005 Paris, France}
\email{jean-philip.piquemal@sorbonne-universite.fr}

\title{The Q-AMOEBA (CF) Polarizable Potential}


\begin{document}

\begin{abstract}
We present Q-AMOEBA (CF), an enhanced version of the Q-AMOEBA polarizable model that integrates a geometry-dependent charge flux (CF) term while designed for an explicit treatment of nuclear quantum effects (NQE). The inclusion of CF effects allows matching experimental data for the molecular structure of water in both gas and liquid phases, addressing limitations faced by most force fields. We show that Q-AMOEBA (CF) provides highly accurate results for a wide range of thermodynamical properties of liquid water. Using the computational efficiency of the adaptive Quantum Thermal Bath method, which accounts for NQE at a cost comparable to classical molecular dynamics, we evaluate the robustness and transferability of Q-AMOEBA (CF) by calculating hydration free energies of various ions and organic molecules. Finally, we apply this methodology to the alanine dipeptide and compute the corresponding dihedral angle potential of mean force and hydration free energy. Unexpectedly, the latter quantity displays significant NQE. These results pave the way to a finer understanding of their role in biochemical systems.

\end{abstract}


The pursuit of a realistic and accurate description of water across its diverse phases has long been central to computational molecular science, particularly in Monte Carlo (MC) and Molecular Dynamics (MD) simulations. 
Since the pioneering efforts of Stillinger and Rahman \cite{stillinger1972molecular,stillinger1974improved}, numerous computational water models have been developed \cite{guillot2002reappraisal,vega2011simulating,kadaoluwa2021systematic,mark2001structure}, each aiming to capture water's intricate properties, such as its permanent dipole moment, non-monotonic density profile, and temperature-dependent dielectric constant to only name a few \cite{nilsson2015structural,stillinger1980water}.
Early models, such as AMBER \cite{case2005amber}, CHARMM \cite{brooks2009charmm}, OPLS \cite{jorgensen1988opls}, SPC \cite{Berendsen1981,berendsen1987missing}, and TIP4P \cite{jorgensen1983comparison,horn2004development,abascal2005general}, employed simplified representations, treating atoms as fixed point charges connected by harmonic bonds. 
While these simple force fields (FF) successfully capture a variety of molecular behaviors, they struggle to accurately represent complex phenomena like polarization and charge transfer, limiting their transferability to more complex environments.
To overcome these limitations, polarizable FF\cite{reviewpol1,reviewpol2,reviewpol3} such as TTMx-F \cite{burnham2002development,fanourgakis2008development}, AMOEBA \cite{ren2003polarizable,ren2004temperature,ponder2010current}, AMOEBA+ \cite{liu2019amoeba+,liu2019implementation}, SIBFA \cite{naseem2022development}, GEM* \cite{duke2014gem}, HIPPO \cite{rackers2021polarizable}, MASTIFF \cite{van2016beyond,van2018new}, CMM~\cite{heindel2024completely}, and others \cite{reviewwater},  were developed, introducing atomic polarizability to account for many-body effects. 
This enhancement allowed for more accurate simulations, particularly in systems characterized by strong electrostatic interactions, such as proteins and ionic liquids. 
However, these models are primarily parametrized for classical MD, and generally overlook the complexities introduced by Nuclear Quantum Effects (NQE), that can prove significant in systems involving light elements like hydrogen \cite{miller2006including,paesani2006accurate,ceriotti2016nuclear}.

Some recent force fields, such as MB-pol \cite{medders2014development,reddy2016accuracy}, q-AQUA-pol~\cite{qu2023interfacing},  ARROW \cite{kurnikov2024neural}; and neural network potentials such as FENNIX \cite{ple2023force} and FENNIX-Bio1\cite{FENNIXBIO1}  have addressed this limitation by incorporating NQE via various strategies.
Indeed, the omission of NQE, especially zero-point energy contributions, underestimates molecular vibrations, which impacts thermodynamic and structural properties and can lead to important discrepancies with experimental results \cite{ceriotti2013nuclear,wilkins2017nuclear}. 
In biological systems, where hydrogen plays a crucial role, addressing NQE is even more critical \cite{agarwal2002nuclear,wang2014quantum,pereyaslavets2018importance}. 
NQE are often considered to be included implicitly in the potential energy surface (PES) of the most common FF, as their parameter values are fitted to reproduce experimental observables when performing MD simulations with classical nuclei. 
In this case, explicit incorporation of NQE requires a prior re-parameterization of the FF \cite{mei2015numerical} to avoid double-counting of the quantum effects. 
NQE can be introduced explicitly via Path Integral Molecular Dynamics (PIMD): based on Feynman's path integral formulation of quantum mechanics \cite{feynman2010quantum}, 
PIMD maps quantum particles to a cyclic chain of classical particles (often called beads), making the method significantly more computationally demanding than classical MD. 
Consequently, PIMD has seen limited use, although several water potentials have been developed using this method \cite{fanourgakis2006quantitative,paesani2007quantum,mcbride2009quantum,habershon2009competing,pereyaslavets2022accurate,zhu2023mb}. 
Different approaches have been proposed to lower PIMD's computational overhead by reducing the required number of beads\cite{kapil2016high,poltavsky2016modeling, markland2008efficient,markland2008refined, ceriotti_PRL2012_PIGLET, brieuc_JCTC2016_PIQTB}, but for most applications, the cost generally remains significantly higher than that of classical MD simulations.
As an alternative to PIMD, the Quantum Thermal Bath (QTB) method \cite{dammak2009quantum,ceriotti2009nuclear} approximates the quantum statistical distribution with a computational cost comparable to classical MD. 
Though the original approach suffers from zero-point energy leakage, limiting its accuracy, a solution was found with the adaptive QTB (adQTB), that enforces the quantum fluctuation-dissipation theorem in a systematic way\cite{mangaud2019fluctuation}.
This method has proven effective in computing NQE in liquid water \cite{mauger2021nuclear} and developing a refined model based on the polarizable AMOEBA PES, named Q-AMOEBA \cite{mauger2022improving}. 
While Q-AMOEBA was shown to provide a remarkably accurate description of various properties of liquid water\cite{mauger2022improving}, its charge distribution does not fully respond to local geometry changes, as it does not account for intramolecular charge transfer or Charge Flux (CF) \cite{palmo2006inclusion,mannfors2008spectroscopically,liu2019implementation}, essential for accurate water modeling \cite{sedghamiz2017probing,sedghamiz2018evaluating}.
Furthermore, previous polarizable models that include a CF term\cite{liu2019amoeba+} were not designed to explicitly account for NQEs.

Therefore, in this paper, we introduce a geometry-dependent CF term into our previously developed Q-AMOEBA model. The new model is referred to as Q-AMOEBA (CF) and was implemented within the GPU-accelerated Tinker-HP software \cite{lagardere2018tinker,adjoua2021tinker} and leverages its Quantum-HP NQE scalable package \cite{ple2023routine}. 
This enhancement seeks to improve the accuracy and transferability of Q-AMOEBA across diverse environments (gas phase, bulk systems, and interfaces).  
We will outline our parametrization procedure, and compare structural and thermodynamic properties of liquid water with the original AMOEBA-03 model \cite{ren2003polarizable}, before exploring the hydration free energies (HFE) of different ions and small molecules.
Finally, to further demonstrate the robustness of the new Q-AMOEBA (CF) model, we turn to a larger compound, the alanine dipeptide, for which we compute the HFE and the dihedral angle potential of mean force, taking advantage of the efficient enhanced sampling methods implemented in Tinker-HP.

The AMOEBA potential energy surface is composed of bonded and nonbonded energy term can be expressed as the sum of bonded and non-bonded energy terms \cite{ren2003polarizable,ponder2010current,liu2019amoeba+}:
\begin{equation}
    \begin{split}
   & E_{total}=E_{bonded}+E_{nonbonded} \\
   & E_{bonded}=E_{bond}+ E_{angle}+E_{b\theta}+E_{CF} \\
   & E_{nonbonded}=E_{vdW}+E_{ele}^{perm}+E_{ele}^{ind}
    \end{split}
\end{equation}
Compared to the previous Q-AMOEBA model \cite{mauger2022improving} (and to the original AMOEBA-03), different parameters were modified. First, the inclusion of the Charge Flux (CF) correction, adapted from the AMOEBA+ model~\cite{liu2019implementation}, which reproduces intramolecular charge reorganization with geometry deformations. In particular, this correction was shown to improve the agreement with the molecular bond angle experimental values across the gas and liquid phases, thereby addressing a common issue in water force fields \cite{ren2003polarizable, liu2019amoeba+, liu2019implementation}. The CF parameters were initialized from the AMOEBA+ parameters \cite{liu2019implementation} and further adjusted to recover the experimental bond length and angle in simulations with explicit treatment of NQE.
Compared to AMOEBA+, only the angular CF parameter (j$_{\Theta}$) was significantly modified, increasing by an order of magnitude, from 0.0020 to 0.0433 e.degree, to account for the enhanced angular fluctuations induced by NQE, while the bond CF parameters remained unchanged (see Table II of SI).
Then, the bond stretching force parameter was slightly reduced to improve the agreement of the simulated infrared (IR) absorption spectra with experimental data. 
Finally, the van der Waals (vdW) parameters were fine-tuned using as target property the experimental density at four temperatures (249.15 K, 277.15 K, 298.15 K, and 341.15 K). The final list of parameters is provided in supporting information (SI).
As discussed below, the incorporation of the CF term and the subsequent adjustments enable the use of a consistent set of intramolecular parameters across gas and liquid phases, with a good match to the experimental data for molecular structure properties (see Table~\ref{parameters}).

In addition, since adQTB and PIMD simulations lead to slightly different values for the liquid water density\cite{mauger2021nuclear, mauger2022improving}, two distinct vdW parameter sets were developed for each of the two methods. 
Unless stated otherwise, the results provided here for the Q-AMOEBA (CF) model are obtained from adQTB simulations (using the adQTB parameters set). The PIMD results are generally very similar, as shown from the detailed analysis available in SI. The SI also provides an extensive comparison with an updated version of our previous Q-AMOEBA model (with reduced bond stretching force parameter but without CF term).

Liquid water simulations used 4000 molecules in a cubic box with periodic boundary conditions. The vdW cutoff was 12 Å, electrostatic interactions were computed with the Smooth Particle Mesh Ewald (SPME) method \cite{essmann1995smooth,SPMEpol} using a 7 Å real-space cutoff and a 60 × 60 × 60 grid. 
A BAOAB-RESPA integrator \cite{Pushing19} with 0.2 fs and 2 fs time steps for bonded and nonbonded interactions, respectively, was used. 
PIMD simulations employed 32 beads with the Thermostated Ring Polymer Molecular Dynamics (TRPMD) method with a mild Langevin thermostat on the centroid with a friction coefficient of  $\gamma = 1$ ps$^{-1}$.
In contrast, the adQTB used a higher friction coefficient of 20 ps$^{-1}$.

\begin{table}[]
\begin{tabular}{ccc|cc}
\hline \hline \TBstrut 
                                     & \multicolumn{2}{c|}{Gas Phase}                                                    & \multicolumn{2}{c}{Liquid Phase}                                                \TBstrut  \\ \cline{2-5} 
                                     & \multicolumn{1}{c|}{$r_{OH}^{eq}$ (\AA)} & $\theta_{HOH}^{eq}$ (deg) & \multicolumn{1}{c|}{$\langle r_{OH}\rangle$ (\AA)} & $\langle\theta_{HOH}\rangle$ (deg) \TBstrut \\ \hline
\multicolumn{1}{c|}{Q-AMOEBA (CF)}   & \multicolumn{1}{c|}{0.957}                                 & 104.50               & \multicolumn{1}{c|}{0.983}                         & 105.15    \TBstrut \\
\multicolumn{1}{c|}{AMOEBA-03}          & \multicolumn{1}{c|}{0.957}                                 & 108.50               & \multicolumn{1}{c|}{0.968}                                 & 106.10             \TBstrut  \\ \hline
\multicolumn{1}{c|}{Exp.}            & \multicolumn{1}{c|}{0.957}                                 & 104.52               & \multicolumn{1}{c|}{0.97}                                  & 105.1               \TBstrut\\ \hline \hline
\end{tabular}
\caption{\label{parameters} Gas-phase equilibrium O-H bond length in \AA\, and H-O-H angle parameters in degree and liquid-phase average values obtained in adQTB molecular dynamics simulations at 300K with Q-AMOEBA (CF) compared to the original AMOEBA-03  model (using classical MD) and to experimental results \cite{SOPER2000121,doi:10.1063/1.1742731}.}
\end{table}

\subsection{Binding energies of Water Clusters}
Table \ref{BE of small Water Clusters} shows the Binding Energies (BE) for various water clusters, comparing the Q-AMOEBA (CF) model with AMOEBA-03 and \textit{ab initio} methods \cite{SOPER2000121,tschumper2002anchoring,smith1990transition}.

This allows us to validate the new parameters against the original AMOEBA-03 and \textit{ab initio} BE. 
The inclusion of CF has reduced the Root Mean Squared Error (RMSE) to 0.37, compared to 0.81 for AMOEBA-03.
\begin{table*}
    \centering
    $
    \begin{array}{*{12}{c}}
    \hline \hline     
    \text{\small{(H$_2$O)$_2$}}     & \text{\small{CCSD(T)} }    &     \text{\small{AMOEBA-03}}   & \text{\small{Q-AMOEBA (CF)} } \TBstrut  \\
    \hline
    \text{\small{Smith01}}     & \text{\small{-4.97} }    &     \text{\small{-4.58}}   & \text{\small{-5.22}}\TBstrut  \\
    \text{\small{Smith02}}     & \text{\small{-4.45} }    &     \text{\small{-3.98}}   & \text{\small{-4.62}} \TBstrut  \\
    \text{\small{Smith03}}     & \text{\small{-4.42} }    &     \text{\small{-3.94}}   & \text{\small{-4.57}}  \TBstrut  \\
    \text{\small{Smith04}}     & \text{\small{-4.25} }    &     \text{\small{-3.54}}   & \text{\small{-3.77}}  \TBstrut  \\
    \text{\small{Smith05}}     & \text{\small{-4.00} }    &     \text{\small{-2.69}}   & \text{\small{-3.35}} \TBstrut  \\
    \text{\small{Smith06}}     & \text{\small{-3.96} }    &     \text{\small{-2.59}}   & \text{\small{-3.25}} \TBstrut  \\
    \text{\small{Smith07}}     & \text{\small{-3.26} }    &     \text{\small{-2.55}}   & \text{\small{-3.05}} \TBstrut  \\
    \text{\small{Smith08}}     & \text{\small{-1.30} }    &     \text{\small{-0.8}}   & \text{\small{-1.26}}  \TBstrut  \\
    \text{\small{Smith09}}     & \text{\small{-3.05} }    &     \text{\small{-2.69}}  & \text{\small{-3.21}} \TBstrut  \\
    \text{\small{Smith10}}     & \text{\small{-2.18} }    &     \text{\small{-1.89}}  & \text{\small{-2.36}} \TBstrut  \\
     \text{\small{RMSE}}     & \text{\small{ } }    &     \text{\small{0.81}}    & \text{\small{0.37 (0.37)}} \TBstrut  \\
    \hline \hline
\end{array}
$
\caption{\label{BE of small Water Clusters}Binding Energies for the 10 Smith dimers with Q-AMOEBA (CF) compared to AMOEBA-03 and \textit{ab initio} references. RMSE values in parentheses are from parameters optimized through PIMD. CCSD(T)/CBS results from \cite{tschumper2002anchoring}.}
\end{table*}
Table IV in SI demonstrates that Q-AMOEBA (CF) reduces the RMSE by about 25\% compared to AMOEBA-03 for larger water clusters, regardless of the method (PIMD or adQTB) used for including NQE during the parametrization. 
This highlights the importance of precise monomer geometry in a flexible water model for faithfully reproducing the intricate PES of water clusters. 

\subsection{Properties of liquid water}
In bulk liquid water, the H$_2$O molecules' structural dynamics generate a globally balanced and symmetric environment, resulting in most molecules having an average net charge close to zero.

Therefore, Q-AMOEBA (CF) only considers intramolecular charge transfers, which are described by the Charge Flux term. 
This CF term couples to the many-body polarization of the system and allows to capture the variations of the molecular dipole with its chemical environment. As a result of this improved description of polarization effects, Q-AMOEBA (CF) predicts a change in dipole moment values from 1.84 D in the gas phase to 2.77 D in the liquid phase.
These results align closely with both \textit{ab initio} MD calculations \cite{silvestrelli1999water} and experimental observations \cite{clough1973dipole,badyal2000electron}, which report a gas-phase dipole moment of 1.85 D and a liquid-phase near 2.9 D. 
In contrast, AMOEBA-03 predicts a similar value for the liquid phase (2.75~D) but underestimates the gas-phase dipole (1.77~D).

More generally, Table~\ref{parameters} presents the properties of intramolecular water structures and shows that the Q-AMOEBA (CF), with its geometry-dependent term, aligns with experimental data across gas and liquid phases.

\begin{figure*}
    \includegraphics[scale=0.73]{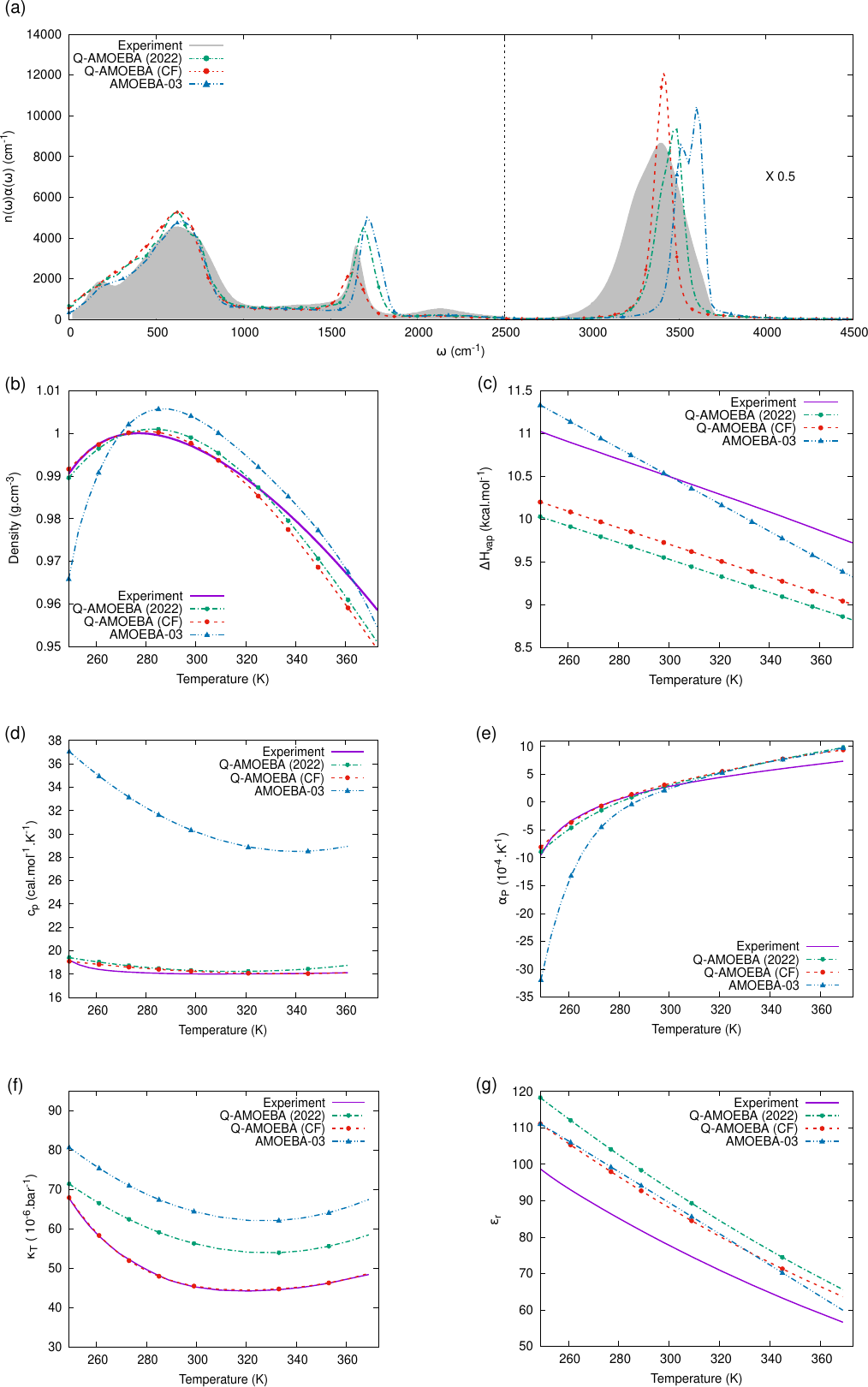}
    \caption{\label{all curves} Liquid properties of water at 1 atm pressure with Q-AMOEBA (CF) compared to AMOEBA-03, previously published Q-AMOEBA (2022) \cite{mauger2022improving} and experimental data \cite{Bertie1989,Kell1975,doi:10.1063/1.1461829,kell1975density}. (a) is the IR absorption spectra computed at 300~K and $\rho=0.997$ g.cm$^{-3}$. The right part of the plot, corresponding to the stretching mode region, is multiplied by 0.5. Various liquid properties of water at a broad range of temperatures and 1 atm pressure: density (b), enthalpy of vaporization (c), isobaric heat capacity (d), thermal expansion coefficient (e), isothermal compressibility (f) and dielectric constant (g). Corresponding radial distribution functions are available in SI.}
\end{figure*}

NQE also significantly influence vibrational spectroscopy, affecting peak positions and intensities in IR spectra. 
Simulations not accounting for these effects often deviate from experimental results. 
Figure \ref{all curves}-(a) shows IR spectra for Q-AMOEBA (CF) with a deconvolution procedure applied to correct for the spectral broadening caused by the high friction coefficient used in adQTB simulations, as previously reported \cite{mauger2021nuclear}. 
The explicit inclusion of NQE generally tends to red-shift the intramolecular peak frequencies\cite{ceriotti2016nuclear}, which is consistent with the results of Fig.~\ref{all curves}-(a), comparing Q-AMOEBA (CF) and AMOEBA-03. This trend is independent on the method used for the inclusion of NQE and the TRPMD curves provided in SI-Fig.4 only display marginal discrepancies with respect to the adQTB results of Fig.~\ref{all curves}-(a). 
Using a lower gas-phase bond stretching force in our updated FF leads to a stretching peak position (at $\sim$3400~cm$^{-1}$) that closely match experimental data, without affecting liquid thermodynamic properties. Additionally, the CF term induces a notable red shift and intensity reduction of the angle bending peak (at $\sim$1600~cm$^{-1}$) , aligning with the experimental spectrum \cite{ramesh2008charge, thompson2000frequency, robertson2002isolating}, and in agreement with previous observations for the AMOEBA+ \cite{liu2019amoeba+} force field.  

Thermodynamic properties such as density, heat capacity, thermal expansion, and isothermal compressibility were assessed from 249.15 K to 369.15 K at 1 atm (see Figure \ref{all curves}-(b)-(g)). 
Q-AMOEBA (CF) accurately reproduces the unusual bell-shaped density curve of water, while AMOEBA-03 shows a notable decrease in density at low temperatures, deviating from experimental observations.
NQE influence water's density maximum due to oxygen's electronegativity, which creates polarized OH bonds and directional hydrogen bonding that stabilizes local tetrahedral structures (similar to that found in ice).
The observed density maximum shifts for different isotopes (277.13K for H$_2$O, 284.34~K for D$_2$O, and 286.55~K for T$_2$O) \cite{jancso1974condensed, ceriotti2016nuclear} indicate an enhanced hydrogen bond network with increasing isotope mass.
Notably, the inclusion of NQE and CF brings the predicted maximum density closer to the experimental value, with Q-AMOEBA (CF) at 279.5 K compared to AMOEBA-03's higher value of 287 K.

The enthalpy of vaporization ($\Delta \text{H}_{\text{vap}}$) is an other characteristic quantity that is known to be strongly affected by NQE in water.
Q-AMOEBA (CF) achieves chemical accuracy for $\Delta \text{H}_{\text{vap}}$ across all temperatures, with a slight temperature-independent shift of $\sim0.8$~kcal.mol$^{-1}$, thereby improving over the previous Q-AMOEBA model (2022) \cite{mauger2022improving}. Though the absolute error of AMOEBA-03 (in classical MD simulations) is lower than that of Q-AMOEBA (CF), this is likely due to a compensation of errors since AMOEBA-03 fails to capture the temperature-dependence of the enthalpy of vaporization and exhibits a strongly overestimated slope. Indeed, it was shown that explicit inclusion of NQE was essential to capture the physical processes at play and recover the correct temperature slope\cite{mauger2022improving}.     

Notably, the enthalpy of vaporization was not used as a fitting property in the parametrization, yet the combined inclusion of CF and NQE reduces the error to within 1 kcal.mol$^{-1}$, while reliably reproducing the experimental temperature dependence of this NQE-sensitive thermodynamic property.

The impact of the CF term on the estimation of the dielectric constant is shown in Figure \ref{all curves}-(g). As discussed above, CF contributes to a more accurate description of the molecular charge distribution, leading to a more accurate dipole moment calculation and a better treatment of polarization effects as a function of molecular geometry.
While Q-AMOEBA (CF) yields a slight improvement over the original Q-AMOEBA FF in estimating the dielectric constant, a noticeable discrepancy with the experimental value remains, indicating that further refinement may be necessary. particular property.

Finally, Q-AMOEBA (CF) predicts a self-diffusion coefficient of $2.42\pm0.01 \times 10^{5}$ cm$^2$/s at 298.15 K, in good agreement with experiments. In comparison, AMOEBA-03’s value of 2.02 × 10$^{5}$ cm$^2$/s demonstrates that including NQE accelerates diffusion compared to classical dynamics. Note that the diffusion coefficient was calculated for Q-AMOEBA (CF) using thermostated ring-polymer MD (TRPMD) as the adQTB method was shown in previous studies\cite{mauger2021nuclear,mauger2022improving} to underestimate diffusion due to the high friction coefficients required to control zero-point energy leakage (the adQTB value for the coefficient is indeed 0.80 ± 0.01 × 10$^{5}$ cm$^2$/s).
Comparisons with Q-AMOEBA (2024), AMOEBA+, and AMOEBA-03 are included in SI to isolate the effects of NQE and CF individually.
Additional studies on ice I$_\text{c}$ and heavy water expand on the role of NQE with Q-AMOEBA (CF) is also available.

\subsection{Hydration Free Energy}
\begin{table}[]
\centering
\begin{tabular}{c|ccccccccc}
\hline \hline
                   & H$_2$O & F$^-$  & Na$^+$  & Mg$^{2+}$ & Cl$^-$  & K$^+$  & Ca$^{2+}$ & Br$^-$ & I$^-$  \TBstrut \\ \hline 
Q-AMOEBA (CF)        & -5.79 (-5.69)  & -116.1 &  -88.5   & -434.4   & -85.7  & -71.4  & -355.0    & -79.6  & -70.8    \TBstrut   \\
AMOEBA-03             & - 5.78  & -114.7 &  -89.7  &  -432.4   &  -84.1  & -72.3  & -354.6    & -77.7  & -69.3  \TBstrut  \\ \cline{1-10} 
Expt.              & -6.32  & -119.7 &  -88.7  &  -435     & -89.1   & -71.2  & -356.8    & -82.7  & -74.3  \TBstrut   \\ \hline \hline
\end{tabular}
\caption{\label{HFE ions}Hydration free energies of water and different ions given in kcal.mol$^{-1}$ obtained with Q-AMOEBA (CF) models compared to AMOEBA-03 and experiments \cite{A907160A}.
The value in parenthesis for water is the one obtained with the parameter set optimized with the PIMD method.
1.9 kcal.mol$^{-1}$ was added to each of the simulated ion hydration hydration free energies. \cite{grossfield2003ion}}
\end{table}

Table \ref{HFE ions} shows the Hydration Free Energies (HFE) computed with Q-AMOEBA (CF) and AMOEBA-03 compared to experimental results for different ions and for water.
The HFE were computed using a free energy perturbation procedure based on twenty-one thermodynamic states in the \textit{NVT} ensemble at 298.15 K.
The detailed MD window parameter can be found in SI.
Each thermodynamical state was thermalized for 1.5 ns and run for 5 ns.
The Bennett Acceptance ratio (BAR) method \cite{bennett1976efficient} and the Path Integral BAR (PI-BAR) for PIMD results \cite{ple2023routine}  was then used to estimate the free energy. 
The final free energy was taken as the sum over all windows, where 1.9 kcal.mol$^{-1}$ was added for each ions \cite{grossfield2003ion}.
For water, the Q-AMOEBA (CF) model with explicit inclusion of NQE yields results that are very close to that of the original AMOEBA-03 (using classical MD), with a slight underestimation of the corresponding HFE with respect to experiment. 

Table~\ref{HFE ions}, shows a clear overall improvement in the HFE values for the different ions. Note that for monovalent ions, this improvement was obtained with the original AMOEBA parameters\cite{grossfield2003ion}, while the Thole parameters have been slightly adjusted for Mg$^{2+}$ and Ca$^{2+}$ to correctly account for NQE with Q-AMOEBA (CF). The new parameters are available in SI.

\begin{figure}
    \centering
    \includegraphics[scale=0.23]{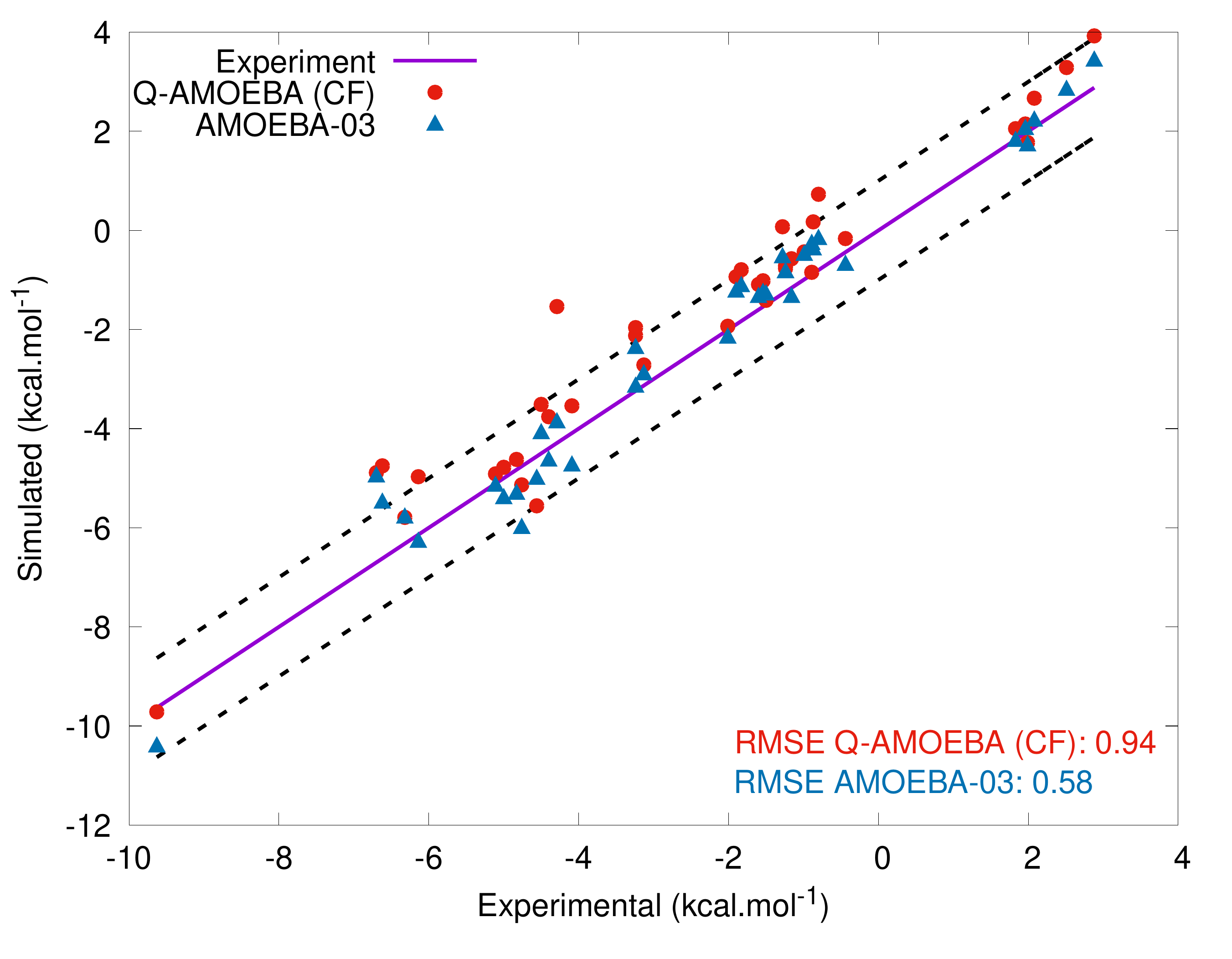}
    \centering
      \caption{Hydration free energies of small organic molecules using Q-AMOEBA (CF) and AMOEBA-03 with their respective RMSE compared to experimental values. The dashed-lines represents the chemical accuracy.}
    \label{HFE molecules}
\end{figure}

To further assess the transferability of Q-AMOEBA (CF), we also proceed to compute the HFE of approximately 40 small organic compounds which were parametrized using Poltype2 \cite{https://doi.org/10.1002/jcc.26954-d}
(All parameter used for the solutes are available in the SI).
Figure \ref{HFE molecules} displays the scatter plot of Q-AMOEBA (CF) models with AMOEBA-03 against experimental values alongside their respective RMSE. 
The original AMOEBA-03 model exhibits better agreement with experimental data (0.58 for AMOEBA-03 and 0.94 for Q-AMOEBA (CF) model).
Despite this, the Q-AMOEBA (CF) model has a strong correlation coefficient $r^2$ (0.97), albeit with slightly higher RMSE values.
Considering the explicit inclusion of NQE in Q-AMOEBA (CF), the higher RMSE values are reasonable, as these effects introduce additional complexity to the model.
Furthermore, none of the solute parameters derived using PolType2 were parameterized with explicit inclusion of NQE. Therefore, the lower RMSE observed with AMOEBA-03 is likely due to a compensation of errors. Further work, focused on the role (and specific parametrization?) of the van der Waals contribution will allow to deepen our understanding of the role of NQE in hydration free energies. 

\subsection{The Alanine Dipeptide}
\begin{figure}
    \centering
    \includegraphics[scale=0.50]{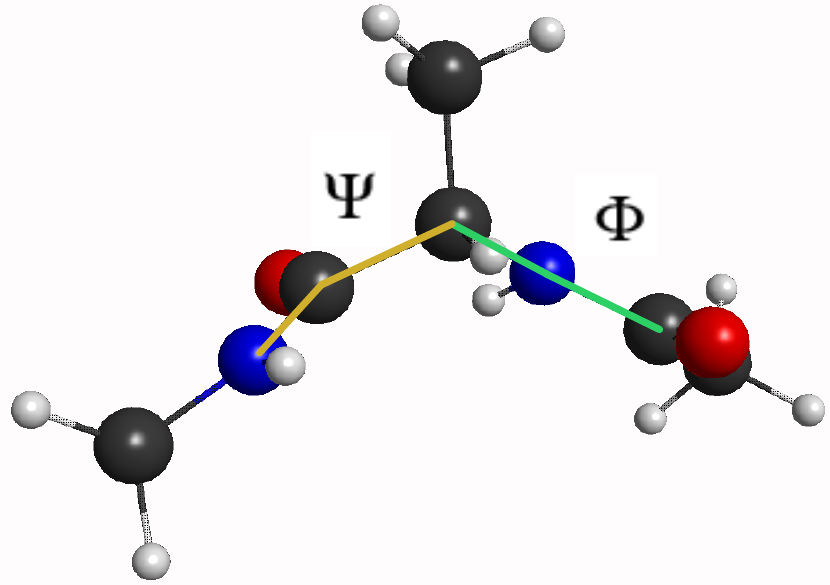}
    \centering
      \caption{Schematic representation of the alanine dipeptide conformation, highlighting the $\Psi$ and $\Phi$ dihedral angles.}
    \label{alanine}
\end{figure}
A longstanding question in the field of biomolecular simulation concerns the quantitative and qualitative impacts of NQE \cite{markland2018nuclear} in proteins and in protein-ligand systems. Such calculations present inherent challenges due to the high number of conformational states accessible.
In this context, we focus on a prototypical system, the alanine dipeptide, a relatively simple yet insightful model that captures key sampling challenges also present in larger proteins.
Indeed, it embodies sampling difficulties typical of larger proteins and its well known metastable states characterized by the two dihedral angles $\Phi$ and $\Psi$ (shown in Figure \ref{alanine}) serves as an ideal model for exploring NQE impacts.

To overcome these, we first resort to the Adaptive Biasing force (ABF) technique \cite{10.1063/1.2829861,comer2015adaptive} using the Colvars library \cite{doi:10.1080/00268976.2013.813594} with both the $\Phi$ and $\Psi$ as collective variables in order to recover the 2 dimensional free energy surface as a function of ($\Phi$,$\Psi$) with NQE using adQTB and without NQE.
Details of the simulation, and input files are provided in SI.
Our simulations (available in SI) reveal only marginal differences in the free energy surfaces, suggesting that NQE have a minimal effect in this particular system. 
This highlights the robustness of both the adQTB method and our model in handling complex biological systems.
We then investigate the impact of NQE on the hydration free energy of alanine dipeptide using the Q-AMOEBA (CF) model with NQE and its classical counterpart. The results, shown in Table \ref{final_ABF_values}, are compared with those from the AMOEBA-03 force field.
To compute this, we leverage the recently introduced Lambda-ABF technique within the Tinker-HP framework, coupled to Colvars \cite{,doi:10.1021/acs.jctc.3c01249}. 
This framework allows for efficient sampling of the alchemical variable $\lambda$ while biasing other slow degrees of freedom, such as $\Phi$ and $\Psi$. 
Given that $\Phi$ is known to be associated with the largest free energy barrier, we perform 2D ABF simulations with respect to both $\lambda$ and $\Phi$, yielding 2D free energy surfaces. 
The hydration free energy of alanine dipeptide is then recovered by marginalizing with respect to $\lambda$.
The hydration free energy computed with the Q-AMOEBA (CF) model and NQE is approximately 4 kcal/mol lower than that obtained using the classical version of the model, indicating that NQE reduces the hydrophilic character of the alanine dipeptide.
While electrostatic contributions remain consistent across all models, vdW interactions vary significantly due to the effects of NQE (Figure available in SI). 
This variation stems from quantum-induced atomic delocalization, particularly in light atoms such as hydrogen, which broadens the range of accessible molecular configurations and enhances entropic contributions. 
As a result, the strength of vdW interactions changes, while electrostatic interactions—primarily determined by fixed charge distributions and distances—are less influenced. 
Ultimately, the enhanced sampling enabled by NQE leads to greater entropic contributions to the hydration free energy, causing discrepancies between the quantum and classical models. 
Interestingly, the AMOEBA-03 results closely matched those of the classical Q-AMOEBA (CF) model, indicating that neglecting NQE, especially in systems where vdW forces play a significant role, can lead to inaccurate free energy estimates.

\begin{table}[h]
\begin{tabular}{c|c}
\hline \hline
                            & Hydration Free Energy \TBstrut\\ \hline
Q-AMOEBA (CF) - adQTB       & -8.37                  \TBstrut\\

Q-AMOEBA (CF) - Classical   & -13.50                  \TBstrut \\
AMOEBA-03                   & -12.25                  \TBstrut\\ \hline \hline
\end{tabular}
 \caption{\label{final_ABF_values}Hydration free energies of the Alanine dipeptide in kcal.mol$^{-1}$ using Lambda-ABF method. Q-AMOEBA (CF) - Classical refers to the results obtained using this FF but with classical dynamics.}
\end{table}
\newpage

In this work, we introduced Q-AMOEBA (CF), an enhanced version of the Q-AMOEBA model that incorporates a geometry-dependent CF term and explicitly accounts for NQE via the adQTB method.
The CF correction addresses key limitations of traditional FF by refining molecular geometries to align bond lengths and angles with experimental data across both gas and liquid phases.
Q-AMOEBA (CF) is designed to be used in simulations with explicit inclusion of NQE and provides a step forward towards a more accurate description of molecular properties, particularly in the context of water and biological systems. 

This enables the prediction of accurate hydration free energies of organic molecules and the study of more complex systems, such as the alanine dipeptide, overall with improved efficiency and precision, showcasing the broad applicability and versatility of Q-AMOEBA (CF). 
While other models include NQE, their computational expense often limits their use in large systems.
The Q-AMOEBA (CF) model offers a balance between accuracy and computational efficiency, marking a milestone in the development of next-generation polarizable force fields for simulating biological and biochemical systems.
It is also important to note that the vdW interactions in biological systems may require further attention in order to explicitly use NQE at large scale. This parametrization task will be the subject of further research in link with the biological importance of NQE in biology and will be effectively addressed using machine learning techniques, which hold great potential in training models that adapt the vdW parameters for biological contexts\cite{gong24}.

\section*{Supporting Information}
All simulation parameters used in this study, including those for water, ions, and alanine dipeptide, are available in the Supporting Information (SI). The SI also provides the full set of Q-AMOEBA (CF) parameters, along with all input files required for reproducibility and additional figures supporting the results discussed in the main text. Results obtained using TRPMD simulations with Q-AMOEBA (CF), as well as the corresponding water parameters, are provided. In addition, the SI contains the parameters for the Q-AMOEBA (2024) water model (adQTB and TRPMD) together with figures illustrating its performance. Comparative results between Q-AMOEBA (CF), Q-AMOEBA (2024) (TRPMD and adQTB), and the original AMOEBA-03 model are also included. A detailed analysis of ice I$_c$ and heavy water is available. The window parameters used for the hydration free energy calculations, as well as the newly derived Thole parameters for Ca$^{2+}$ and Mg$^{2+}$, are provided. Parameters used for all solutes studied are bundled in a tar archive. Finally, for ease of reproducibility, the Colvar input files used for alanine dipeptide simulations are provided, along with the PMF differences obtained with Q-AMOEBA (CF), Q-AMOEBA (2024) (adQTB and PIMD), and AMOEBA-03, as well as the decomposition of the computed solvation free energy for all models.

\section*{Acknowledgments}

This work has received funding from the European Research Council (ERC) under the European Union's Horizon 2020 research and innovation program (grant agreement No 810367), project EMC2 (JPP). Computations have been performed at IDRIS (Jean Zay) on GENCI Grants: no A0150712052 (J.-P. P.).

\bibliography{acs-achemso}

\end{document}